\documentclass[runningheads,a4paper]{llncs}

\usepackage{amssymb}
\usepackage{graphicx}

\usepackage{url}
\urldef{\mailsa}\path|{jresch, iocelews,jehrentr,mbc}@uwaterloo.ca|    
\newcommand{\keywords}[1]{\par\addvspace\baselineskip
\noindent\keywordname\enspace\ignorespaces#1}

\begin{document}

\mainmatter  

\title{Gamified Automation in Immersive Media for Education and Research}

\titlerunning{Gamified Automation in Immersive Media for Education and Research}

%
%
\author{Janelle Resch%
\and Ireneusz (Eric) Ocelewski
\and Judy Ehrentraut \and\\ Mike Barnett-Cowan}
\authorrunning{Gamified Automation in Immersive Media for Education and Research}

\institute{University of Waterloo, 200 University Ave. W., Waterloo Ontario, N2L 3G1, Canada\\
\mailsa}

%
%

\toctitle{13th International Conference on Persuasive Technology}
\tocauthor{Resch, Ocelewski, Ehrentraut, Barnett-Cowan}
\maketitle

\begin{abstract}
The potential of using video games as well as gaming engines for educational and research purposes is promising, especially with the current progress of Industry 4.0 technologies such as augmented and virtual reality devices. However, it is important to be aware of the barriers of these technologies. Integrating additional software libraries into current developer environments would not only increase accessibility for the general public, but would gamify the development process through multiplayer functionality. In this paper we will briefly discuss a couple case studies demonstrating the usefulness of using Unreal Engine 4 for collaborative research purposes. In addition, we present several ideas on how to extend such development environments to further assist researchers and students to easily create prototypes for the purpose of gamified scientific creativity and inquiry. 
\end{abstract}

\section{Introduction and Motivation}

The influence of the \$110 billion and growing videogame industry has been profound on society. More than 2.2 billion people are playing games worldwide and of those, over half play online games \cite{Newzoo}. It is one of the largest growing global markets and with the exponential development of Industry 4.0 technologies, this trend is likely to continue. Traditionally, video games have been created mainly for entertainment purposes. However, the rapid growth of gaming technologies has also been beneficial within academia, but as we argue, is still under appreciated with tremendous unlocked potential for innovation.

For instance, although the original intended use for graphics processing units (GPUs) was to compute and display computer graphics, it is now standard to use GPUs for general purpose computing including scientific computing. This is due to their parallel structure and affordability from being a consumer product. In addition, with the development of platforms such as OpenCL, DirectCompute or NVIDIA’s Compute Unified Device Architecture (CUDA), programmable GPUs are now commonplace technologies for computational mathematics and artificial intelligence (AI). A further outcome of utilizing GPUs as science, technology, engineering and mathematics (STEM) computing devices is that graphics cards can be used as novel educational tools \cite{NvidiaHistory}.

This naturally leads to the concept of \textit{gamification}, which is a term used to describe the use of game design elements and technologies in non-game contexts \cite{Deterding}. The potential of `gamifying' education as well as research and development (R\&D) is significant. One reason for this is due to the intrinsic association with video games: they are enjoyable, social and provide a digital space where people are free to explore, experiment and take risks without severe consequences. Furthermore, video games are a great way to introduce people of all ages to technology and more generally, computer science. Another reason is because of the recent boom in immersive media via mixed reality (MR) technologies such as virtual reality (VR) and augmented reality (AR) devices. Depending on the VR or AR setup however, such technologies (e.g., Oculus Rift, HTC Vive, Microsoft HoloLens, etc.) can be costly due to additional hardware requirements. Fortunately, the cost of GPUs is steadily decreasing and is more accessible  now than ever. Nonetheless, there is a misconception that an expensive setup is required to have VR or AR experiences whereas much of these tools can be optimized for mobile devices.

\begin{figure}
\centering
\includegraphics[scale = 0.35]{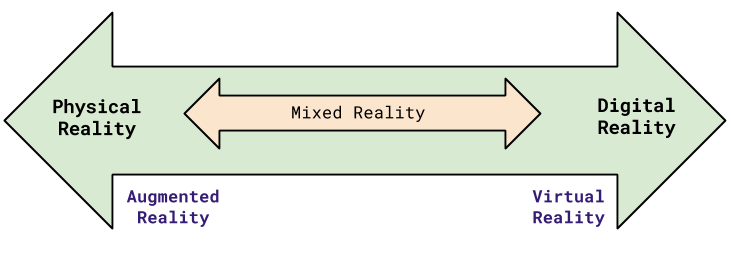}
\caption{The mixed reality spectrum, a concept presented in \cite{Milgram}.}
\label{fi:BP}
\end{figure}

Mixed reality has much potential because it allows for the creation of three-dimensional (3D) interactive learning environments giving users the ability to interact and manipulate the virtual environment. For STEM subject matter, this is especially useful for students, teachers and researchers since it allows one to `play' with a problem. This results in better understanding of the theory and corresponding applications. Further leveraging the software used to develop MR content can also be of great value to students because it teaches them how to approach computation and general STEM problems. Video game engines such an Unreal Engine 4 (UE4) make video game development more accessible to the general public regardless of a person's background. Traditionally, software development (including video games) has required one to specialize in the subject in order to contribute in the industry. However,  there is still a gap into the development of such technologies and the corresponding software. Moreover, there is still a significant intimidating factor and learning curve that cannot be overlooked. One objective of this paper is to present solutions on how to resolve this issue. 

Mixed reality technologies and the corresponding software for the purpose of developing gamified approaches toward education, R\&D, remote training and testing is of great interest to the authors. There exists a hierarchy between software programs, packages and libraries; yet there is still much room for improvement, particularly in multiplayer gamified learning environments or collaboration spaces. It is of the authors’ beliefs that giving users the ability to quickly prototype an idea and then explore in a collaborative manner would be hugely beneficial. For STEM literacy to significantly improve, people need to be shown how a computer can do work for them in an organic way (i.e., with automated components by constructing scripts, leveraging artificial intelligence, neural networks, etc.). Discussing specific toolsets to accomplish this will be another main objective of this paper.

\section{Gamification of Research Case Studies}

Describing and simulating scientific problems can be difficult and time consuming; especially when modeling full three-dimensional (3D) problems. Thus, for the majority of scientific problems, serial computation is not an option and even GPU simulations can take substantial time. This can be a hurdle for educators and researchers since there is not much time for computational experimentation. The expectations within academia are so high that it can become an actual barrier to scientific creativity and progress. We posit that videogame and Industry 4.0 technologies can mitigate this.

Although the physics incorporated in video games does not have the same level of precision as scientific modelling and computing, simulating the physics to first or second order is still faster than faking the science (i.e., using special effect illusions). This means that using such technologies can be a huge benefit for researchers to develop insights on their work, especially when utilizing MR. For example, 3D plotting is useful, but on a 2D screen, it is still only so useful. If one has the option to explore a problem in all three spatial dimensions, it could be of great benefit. 

For instance, one of the authors studies the production and propagation of finite-amplitude sound waves in brass musical instruments, specifically the trumpet and trombone. One aspect the author is currently focused on is understanding and mathematically describing the relationship between pressure and velocity at the mouthpiece of brass instruments when musical notes are being produced by a player. If a musical note being played quietly, it means the amplitude of the sound pressure wave at the mouthpiece of the instrument is a small fraction of atmospheric pressure. Mathematically in such cases, a small amplitude linearization of the equations in gasdynamic can be considered resulting in a linear model of sound propagation. However, for loudly played notes, pressure variations within the narrow bore near the mouthpiece can be a significant fraction of atmospheric pressure. A linearization can no longer be performed since the nonlinear behavior from the high amplitude propagating waves can distort the travelling waves. This physically corresponds to the crest of the wave travelling faster than the trough, i.e., the waveform steepens and can even produce shock waves in certain circumstances. The acoustic consequence of wave steepening is the transfer of energy to the higher frequency components giving the timbre a  more  ‘brassy’  effect. From a modeling perspective, this means that the standard linear relationship between pressure and velocity cannot be imposed as a boundary condition for numerical simulations. However, the nonlinear relationship between pressure and velocity in not known and this is what one of the authors is studying \cite{JR}.

\begin{figure}
\centering
\includegraphics[scale = 0.2]{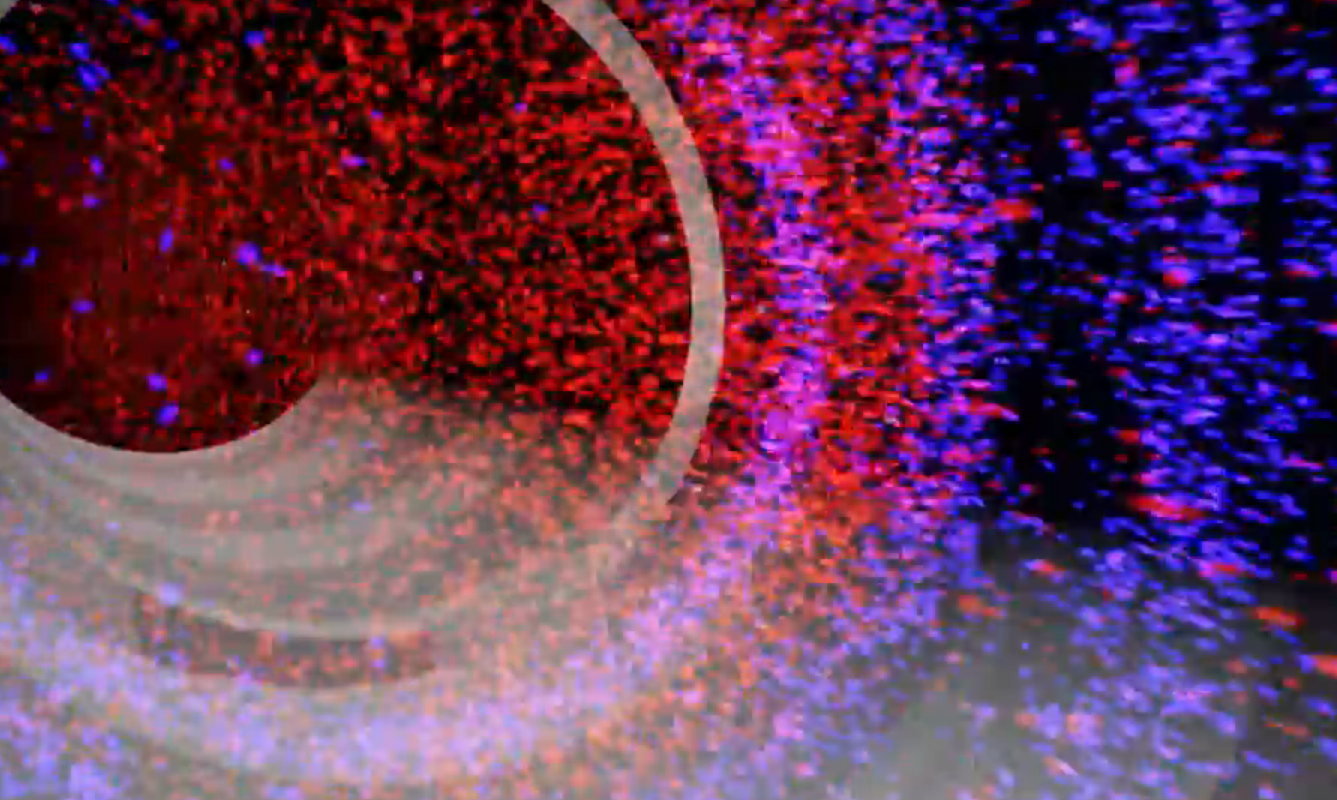}
\caption{Screenshot of the UE4 environment while using NVIDIA libraries to visualize  particles moving through a tube where a step function is considered. The red particles are moving in the forward direction, the blue particles depict which have been reflected from the step in the bore wall.}
\label{fi:UE4Bore}
\end{figure}

To gain inspiration on how to think of the physics of the problem, the author wanted to take advantage of VR. In particular, having the ability to move around (mathematically, this corresponds going from Eulerian to Lagrangian coordinates) in the musical instrument and visualize what is happening at the mouthpiece provides a better understanding of the dynamics of the system. To accomplish this in a timely manner, one of the other authors created a VR environment using UE4 and taking advantage of certain software libraries written by NVIDIA which use accelerated GPU computing. A screenshot of the VR experience is displayed in Figure \ref{fi:UE4Bore} and a brief overview on UE4 and the libraries used is presented in Section \ref{secUE4}.

This example demonstrates how gaming engines with the addition of powerful, scientific software libraries can be paramount as tools for rapid development of scientific inquiry in addition to teaching tools. Another one of the authors recently enabled each student in his class to rapidly develop a demonstration of how MR technology could be used to better understand how the central nervous system integrates multisensory information. Using the Microsoft HoloLens and pairing undergraduate students in kinesiology with little to no background in computer programming with experienced UE4 developers, students were able to focus on their research concepts of potential MR neuroscience experiments while developers focused on rapid development while teaching students how to code. If certain additional aspects were integrated into the engine to help with this learning process, it could evolve into a standard component within academia including teaching methods. This will be further discussed in Section \ref{secEditor}.

\section{The Feasibility of Using Unreal Engine 4 as a Way to Explore Physical Systems} \label{secUE4}

Unreal Engine 4 (UE4) is a video game developer platform from Epic Games Incorporated that can be used to develop VR and AR experiences. The engine is written mostly in C$++$ and uses some C\#. Developers can implement their game environment using C$++$ or they can use the blueprints visual scripting system. Blueprints are constructed with a node-based interface to create gameplay element within the developer environment (called the Unreal Editor) \cite{UE4}. A blueprint in the editor is equivalent to an  object-oriented class or object in C$++$. Within each blueprint, different color nodes are used to define an event, or function, or variable, etc., and are connected with wires each with a specified colour. The color defines the operation or object being implemented. An example of a blueprint in shown in Figure \ref{fi:BP}. When a project uses blueprint, UE4 converts the blueprint into C$++$. Using blueprints is especially useful for quick prototyping of an idea \cite{UE4}.

\begin{figure}
\centering
\includegraphics[scale = 0.35]{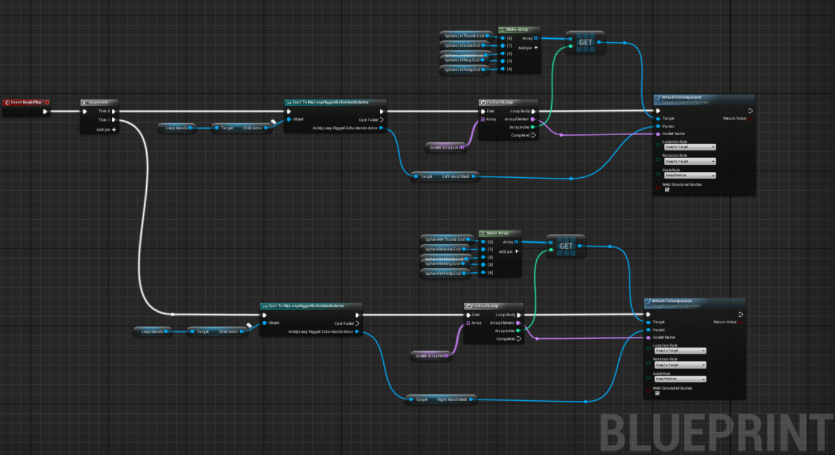}
\caption{Example of a blueprint in UE4. Red nodes are events; blue nodes are functions; purple nodes are socket variables; green nodes are associated with get something (in this case, an array).}
\label{fi:BP}
\end{figure}

Unreal Engine 4 is also a powerful tool because it  accurately simulates the physics of a situation to first order. In particular, UE4 uses PhysX for its physical simulations. NVIDIA FleX and Flow are other GPU software libraries for real-time computational fluid dynamics (CFD) that can be integrated into the UE4 development environment. In particular, FleX is a particle based fluid simulation technique and Flow is a dynamic grid based  gas simulation technique. As NVIDIA describes, stability and speed rather than accuracy are the focus \cite{NVIDIA}, \cite{Flow}, \cite{Flex}. A brief description on these libraries is provided below.

\subsubsection{NVIDIA’s FleX vs Flow:}

\begin{itemize}
\item[] FleX: Usually such particle based simulations result on solvers for fluids and rigid bodies. However, FleX uses particle representations for all types of simulated objects that combines  smoothed particle hydrodynamics (SPH) with position based dynamics (PBD) where larger timesteps can be used as well as smaller smoothing radii. The PBD used is unconditionally stables and uses a nonlinear constraint solver either on distance (mass spring), non-penetration (collision), bending (triangle mesh) or density (i.e., incompressibility assumption) in which the SPH density estimation is defined with respect to a smoothing kernel. To solve the PBD constraints, Newton’s method is applied to each constraint individually and iterated until there is convergence. Finally, FleX simulations have an artificial pressure, so there is no accumulation of pressure for each iteration. This is used to avoid the clumping of particles and it is achieved by adding a small repulsive term to the constraint used \cite{Flex}.

\item [] Flow: Currently available as a beta and does not assume incompressibility. Simulation blocks are allocated dynamically and are set as active if they are areas of interest. The output of the fluid simulation is volume rendered. In the beta, the density and temperature fields are viewed as a colour map with specified opacities where the temperature and density field are ray marched. Flow uses an adaptive sparse voxel which focuses the memory and computation resources on the specific regions of interest (i.e., is a local method) \cite{Flow}.
\end{itemize}

\section{The Editor is the Game} \label{secEditor}

As foreshadowed above, the UE4 editor can be extended by making the editor the game. This would effectively gamify the editor development tools into a social space, thereby creating a social phenomenon that spreads virally. User adoption would depend on access, easy of use and expandability. Current online VR offerings are limited by their infrastructure since development of these platforms is centralized. Using UE4 as a foundational technology for such a platform is ideal since market momentum depends on scale, market penetration and the effective sum total of users who develop or play games powered by UE4. However, it can still be made more user friendly by 1) integrating automation and learning into the user experience (UX) / user interface (UI); and 2) making it multiplayer and shareable. This would also set a more stable foundation for how children will explore and play in the future.

To reduce the learning curve associated with the engine, the authors believe that artificial intelligence (AI) should be integrated into the development environment.  More precisely, the use of AI needs to be gamified in a way that assists the user throughout the learning process; and that ties into a story based narrative that exposes the user to skills necessary for future multiplayer modifications. In addition, these AI could exist and be hosted on an audited, secure network, such as blockchain. This technology would be optimal as it adds and saves metadata to the ledger regularly. We could then model, a leaky integrate-and-fire (LIF) neuron model with populations of neurons and a neural/synaptic network/circuit that interfaces with modules such as computer vision, cleverbot, speech recognition and translation, etc.. Experimenting with various cost functions may be wise, but also setting on unchangeable ones may be preferable. Perhaps two computational universes can be considered and run simultaneously: one with solid rules, another with more flexible rules.

Additionally, incorporating the collection of sensory data able to measure and quantify one's mood would be another key component in this new state-of-the-art engine. Using machine learning algorithms and neural networks to analyze user data would allow for biometric data anomaly prediction and prescriptive automation. This information could provide insight on the physiology of learning and provide customizable approaches to introduce people to new information. Consider for instance that a user begins a session but is grumpy and not in the best mindset. The sensor array could infer the person's mood from reading and analyzing their biometric data, and  then, upon warning (e.g., full disclosure: warning this experience is being modified for your happiness) could procedurally generate and alter experience by narrating an environment, setting, tone, pace or general progression. The question is after studying the data, will the algorithm be able to create a model and teach any user? For this to be a possibility, a playground with a sandbox is needed with a fair playing field and an interface for other players to experiment with AI. 

\section{Conclusion}

In summary, utilizing video game technologies, both in hardware and software can be a game changer for how society progresses into the 21st century. People will need to acquire and maintain information technology (IT) literacy and skills for modern economies. Programing machines will become a requisite skill with coding becoming as ubiquitous as English or Mandarin is today. Leveraging GPU accelerated libraries in UE4 can greatly accelerate the arrival of automation in education and research. This allows for the development of accurate immersive simulations, and virtual environments that can be more STEM oriented with a focus on lifelong learning. 

A strategically constructed framework of play and learning could provide a playground for the researchers of today and the scientist of tomorrow (i.e., children). Without this type gamification, creating experiences for enjoyable learning and discovery may be unnecessarily delayed. Such a space would provide a more social learning environment that has the ability to improve decision making capabilities with the aid of automation and biometric data. Currently, people do work on a computer rather than having a computer do work for them. It needs to be demonstrated how computers can be put to work for humanity. 

\subsubsection*{Acknowledgments.} The authors would like to thank Nathan Zavaglia and Corbin Resch for their input. This work was made possible in part by support from Wellcome Trust, Epic Games, HTC, and NVIDIA (for the Titan Xp). Thank you for the support.


\begin{thebibliography}{4}

\bibitem{Newzoo} McDonald, E.: Newzoo's 2017 Report: Insights Into the \$108.9 Billion Global Games Market. From Global Games Market Report \url{newzoo.com/insights/articles/newzoo-2017-report-insights-into-the-108-9-billion-global-games-market/} (2017).

\bibitem{NvidiaHistory} NVIDIA History, A Timeline of Innovation,\url{www.nvidia.ca/page/corporate_timeline.html}.

\bibitem{Deterding} Deterding, S., Dixon, D., Khaled, R., Nacke, L.: From Game Design Elements to Gamefulness: Defining Gamification. Proceedings of the 15th International Academic MindTrek Conference: Envisioning Future Media Environments (AMC), 9-15(2011).

\bibitem{JR} Resch, J., Krivodonova, L., Vanderkooy, J.: A Two-Dimensional Study of Finite Amplitude Sound Waves in a Trumpet Using the Discontinuous Galerkin Method, Journal of Computational Acoustics, 22(3), 27 pages (2014).

\bibitem{Maitem} Maitem, J., Cabauatan, R. J., Rabago, L., Tanguilig, B.: Math world: A Game-Based 3D Virtual Learning Environment (3D VLE) for Second Graders.The International Journal of Multimedia and Its Applications (IJMA), 4(1), 13 page (2012). 

\bibitem{Toback} Toback, David, Mershin, A., Novikova, I.: Integrating Web-Based Teaching Tools Into Large University Physics Courses. The Physics Teacher, 43(9),  594-597 (2005).


\bibitem{Milgram} Milgram, P., Takemura, H., Utsumi, A., Kishino, F.: Augmented reality: A Class of Displays on the Reality-Virtuality Continuum.  International Society for Optics and Photonics, 2351, 282-293 (1995).

\bibitem{UE4} Unreal Engine 4 Documentation, \url{docs.unrealengine.com/latest/INT}.

\bibitem{NVIDIA} Macklin, M., Muller, M., Chentanez, N., Kim, T.Y.; Unified Particle Physics for Real-Time Application., ACM Transactions on Graphics, 33(4), 12 pages 2014.


\bibitem{Flow} NVIDIA Flow, \url{developer.nvidia.com/nvidia-flow}. 

\bibitem{Flex} NVIDIA FleX,\url{developer.nvidia.com/flex}.


\end{thebibliography}
\end{document}